\documentclass[twocolumn,twoside,slac_two]{revtex4}
\usepackage{graphicx}
\usepackage{fancyhdr}
\def\pr{{Phys. Rev.}~}
\def\prl{{ Phys. Rev. Lett.}~}
\def\pl{{ Phys. Lett.}~}
\def\np{{ Nucl. Phys.}~}

\begin{document}

\title{Standard Model expectations on $\sin2\beta(\phi_1)$ from $b \to s$ penguins}

%

\author{Chun-Khiang Chua}
\affiliation{Institute of Physics, Academia Sinica, Taipei, Taiwan
115, Republic of China}

\begin{abstract}
Recent results of the standard model expectations on
$\sin2\beta_{\rm eff}$ from penguin-dominated $b \to s$ decays are
briefly reviewed.
\end{abstract}

\maketitle

\thispagestyle{fancy}

\section{Introduction}

Although the Standard Model is very successful, New Physics is
called for in various places, such as neutrino-oscillation, dark
matter (energy) and baryon-asymmetry. Possible New Physics beyond
the Standard Model is being intensively searched via the
measurements of time-dependent CP asymmetries in neutral $B$ meson
decays into final CP eigenstates defined by
 \begin{eqnarray}
 &&{\Gamma(\overline B(t)\to f)-\Gamma(B(t)\to f)\over
 \Gamma(\overline B(t)\to f)+\Gamma(B(t)\to f)}
 \nonumber\\
 &&={\cal S}_f\sin(\Delta mt)+{\cal A}_f\cos(\Delta mt),
 \end{eqnarray}
where $\Delta m$ is the mass difference of the two neutral $B$
eigenstates, $S_f$ monitors mixing-induced CP asymmetry and ${\cal
A}_f$ measures direct CP violation. 
The $CP$-violating parameters
${\cal A}_f$ and ${\cal S}_f$ can be expressed as
 \begin{eqnarray}
 {\cal A}_f=-{1-|\lambda_f|^2\over 1+|\lambda_f|^2},
 \qquad
 {\cal S}_f={2\,{\rm Im}\lambda_f\over 1+|\lambda_f|^2},
 \end{eqnarray}
where
 \begin{eqnarray}
 \lambda_f={q_B\over p_B}\,{A(\overline B^0\to f)\over A(B^0\to f)}.
 \end{eqnarray}

\begin{figure}[b]
\centering
\includegraphics[width=80mm]{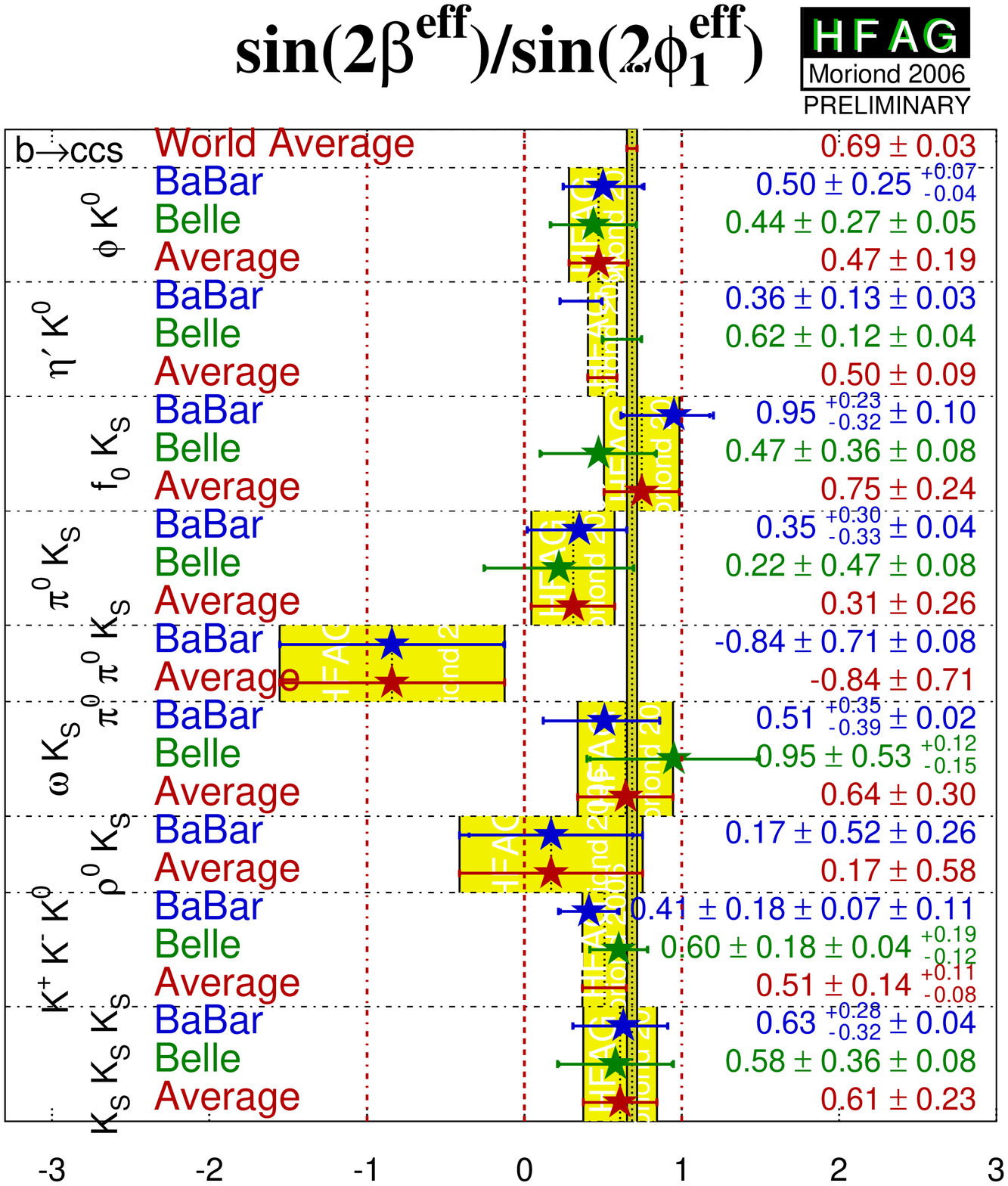}
\caption{Experimental results for $\sin2\beta_{\rm eff}$ from
$b\to s$ penguin decays~\cite{HFAG2006}.} \label{example_figure}
\end{figure}

In the standard model $\lambda_f\approx \eta_f e^{-2i\beta}$ for
$b\to s$ penguin-dominated or pure penguin modes with $\eta_f=1$
($-1$) for final $CP$-even (odd) states. Therefore, it is expected
in the Standard Model that $-\eta_fS_f\approx \sin 2\beta$ and
${\cal A}_f\approx 0$ with $\beta$ being one of the angles of the
unitarity triangle.

The mixing-induced $CP$ violation in $B$ decays has been already
observed in the golden mode $B^0\to J/\psi K_S$ for several years.
The current world average the mixing-induced asymmetry from tree
$b\to c\bar c s$ transition is~\cite{HFAG2006}
 \begin{eqnarray}
 \sin 2\beta=0.687\pm0.032\,.
 \end{eqnarray}
However, the time-dependent {\it CP}-asymmetries in the $b\to
sq\bar q$ induced two-body decays such as $B^0\to
(\phi,\omega,\pi^0,\eta',f_0)K_S$ are found to show some
indications of deviations from the expectation of the Standard
Model (SM)~\cite{HFAG2006} (see Fig.~1). In the SM, CP asymmetry
in all above-mentioned modes should be equal to ${\cal S}_{J/\psi
K}$ with a small deviation {\it at most} ${\cal O}(0.1)$
\cite{LS}. As discussed in \cite{LS}, this may originate from the
${\cal O}(\lambda^2)$ truncation and from the subdominant
(color-suppressed) tree contribution to these processes. Since the
penguin loop contributions are sensitive to high virtuality, New
Physics beyond the SM may contribute to ${\cal S}_f$ through the
heavy particles in the loops. In order to detect the signal of New
Physics unambiguously in the penguin $b\to s$ modes, it is of
great importance to examine how much of the deviation of $S_f$
from ${\cal S}_{J/\psi K}$,
 \begin{eqnarray}
 \Delta {\cal S}_f\equiv -\eta_f {\cal S}_f-{\cal S}_{J/\psi K_S},
 \end{eqnarray}
is allowed in the
SM~\cite{LS,Grossman97,Grossman98,Grossman03,Gronau-piK,Gronau-eta'K,GronauRosner,QCDF,CCY2005,pQCD,SCET,CCS2005,CCSKKK,SU3KKK}.

The decay amplitude for the pure penguin or penguin-dominated
charmless $B$ decay in general has the form
 \begin{eqnarray}
 M(\overline B^0\to f) = V_{ub}V_{us}^*F^u+V_{cb}V_{cs}^* F^c
 +V_{tb}V_{ts}^*F^t.
 \end{eqnarray}
Unitarity of the CKM matrix elements leads to
 \begin{eqnarray}
 M(\overline B^0\to f) &=& V_{ub}V_{us}^*A_f^u+V_{cb}V_{cs}^*A_f^c
 \nonumber\\
 &\approx& A\lambda^4R_be^{-i\gamma}A_f^u+A\lambda^2 A_f^c,
 \end{eqnarray}
where $A_f^u=F^u-F^t$, $A_f^c=F^c-F^t$,
$R_b\equiv|V_{ud}V_{ub}/(V_{cd}V_{cb})|=\sqrt{\bar\rho^2+\bar\eta^2}$.
The first term is suppressed by a factor of $\lambda^2$ relative
to the second term. For a pure penguin decay such as $B^0\to\phi
K^0$, it is naively expected that $A_f^u$ is in general comparable
to $A_f^c$ in magnitude. Therefore, to a good approximation
$-\eta_fS_f\approx\sin 2\beta\approx S_{J/\psi K}$. For
penguin-dominated modes such as $\omega K_S,\rho^0K_S,\pi^0K_S$,
$A_f^u$ also receives tree contributions from the $b\to u\bar u s$
tree operators. Since the Wilson coefficient for the penguin
operator is smaller than the one for the tree operator, $A_f^u$
could be significantly larger than $A_f^c$. As the first term
carries a weak phase $\gamma$, it is possible that $S_f$ is
subject to a significant ``tree pollution". To quantify the
deviation, it is known that to the first order in
$r_f\equiv(\lambda_u A_f^u)/(\lambda_c
A_f^c)$~\cite{Gronau,Grossman03}
 \begin{eqnarray} \label{eq:CfSf}
 \Delta {\cal S}_f=2|r_f|\cos 2\beta\sin\gamma\cos\delta_f,
 \,\, {\cal A}_f=2|r_f|\sin\gamma\sin\delta_f,
 \nonumber
 \end{eqnarray}
with $\delta_f={\rm arg}(A_f^u/A_f^c)$. Hence, the magnitude of
the CP asymmetry difference $\Delta {\cal S}_f$ and direct CP
violation are both governed by the size of $A_f^u/A_f^c$. However,
for the aforementioned penguin-dominated modes, the tree
contribution is color suppressed and hence in practice the
deviation of ${\cal S}_f$ is expected to be small~\cite{LS}. It is
useful to note that $\Delta {\cal S}_f$ is proportional to the
real part of $A^u_f/A^c_f$ as shown in the above equation.

Below I will review the results of the SM expectations on $\Delta
{\cal S}_f$ from short-distance and long-distance calculations.
Recent reviews of results obtained from the SU(3) approach can be
found in~\cite{SU3}.

\section{$\Delta{\cal S}_f$ from short-distance calculations}

There are several QCD-based approaches in calculating hadronic $B$
decays~\cite{BBNS,KLS,BFPS}. $\Delta {\cal S}_f$ from calculations
of QCDF~\cite{QCDF,CCY2005}, pQCD~\cite{pQCD}, SCET~\cite{SCET}
are shown in Table~1. The QCDF calculations on $PP$, $VP$ modes
are from \cite{QCDF}~\footnote{Results obtained agree with those
in \cite{CCS2005}.}, while those in $SP$ modes are from
\cite{CCY2005}. It is interesting to note that (i) $\Delta {\cal
S}_f$ are small and positive in most cases, while experimental
central values for $\Delta {\cal S}_f$ are all negative, except
the one from $f_0 K_S$; (ii) QCDF and pQCD results agree with each
other, since the main difference of these two approach is the
(penguin) annihilation contribution, which hardly affects $S_f$;
(iii) The SCET results involve some non-perturbative contributions
fitted from data. These contributions affect $\Delta S_f$ and give
results in the $\eta' K_S$ mode different from the QCDF ones.

\begin{table}[t]
\begin{center}
\caption{$\Delta {\cal S}_f$ from various short-distance calculations.} 
\begin{tabular}{|l| c c c |c|}
\hline$\Delta {\cal S}_f$
       &QCDF
       &pQCD
       &SCET
       & Expt
       \\
       \hline
 $\phi K_S$
       & $0.02\pm0.01$
       & $0.020^{+0.005}_{-0.008}$
       &
       & $-0.22\pm0.19$
       \\
 $\omega K_S$
       & $0.13\pm0.08$
       &
       &
       & $-0.06\pm{0.30}$
       \\
 $\rho^0K_S$
       & $-0.08^{+0.08}_{-0.12}$
       &
       &
       & $-0.52\pm0.58$
       \\
 $\eta' K_S$
       & $0.01\pm0.01$
       &
       & $\begin{array}{c}-0.02\pm0.01\\-0.01\pm0.01\end{array}$
       & $-0.19\pm0.09$
       \\
 $\eta K_S$
       & $0.10^{+0.11}_{-0.07}$
       &
       & $\begin{array}{c}-0.03\pm0.17\\+0.07\pm0.14\end{array}$
       &
       \\
 $\pi^0K_S$
       & $0.07^{+0.05}_{-0.04}$
       & $0.06^{+0.02}_{-0.03}$
       & $0.08\pm0.03$
       & $-0.38\pm0.26$
       \\
 $f_0K_S$
       & $0.02\pm0.00$
       &
       &  
       & $+0.06\pm0.24$
       \\
  $a_0K_S$
       & $0.02\pm0.01$
       &
       &
       &
       \\
  $\bar K^{*0}_0\pi^0$
       & $\begin{array}{c}0.00^{+0.03}_{-0.05}\\0.02^{+0.00}_{-0.02}\end{array}$
       &
       &
       &
       \\
       \hline
\end{tabular}
\end{center}
\end{table}

\begin{table*}[t]
\begin{center}
\caption{Direct CP asymmetry parameter ${\cal A}_f$ and the
mixing-induced CP parameter $\Delta {\cal S}_f^{SD+LD}$ for
various modes. The first and second theoretical errors correspond
to the SD and LD ones, respectively~\cite{CCS2005}.}
\label{tab:Sf}
\begin{tabular}{|l| r c c |r c c|}
\hline
      &  \multicolumn{3}{c|}{$\Delta {\cal S}_f$}
      &   \multicolumn{3}{c|}{${\cal A}_f(\%)$}
      \\
      \cline{2-4}
      \cline{5-7}
\raisebox{2.0ex}[0cm][0cm]{Final State} & SD & SD+LD & Expt & SD &
SD+LD & Expt
      \\
      \hline
 $\phi K_S$
       & $0.02^{+0.01}_{-0.02}$
       & $0.04^{+0.01+0.01}_{-0.02-0.02}$
       & $-0.22\pm0.19$
      & $0.8^{+0.5}_{-0.2}$
      & $-2.3^{+0.9+2.2}_{-1.0-5.1}$
      & $9\pm14$
      \\
 $\omega K_S$
       & $0.12^{+0.06}_{-0.05}$
       & $0.02^{+0.03+0.03}_{-0.04-0.02}$
       & $-0.06\pm{0.30}$
      & $-6.8^{+2.4}_{-4.0}$
      & $-13.5^{+3.5+2.4}_{-5.7-1.5}$
      & $44\pm23$
      \\
 $\rho^0K_S$
       & $-0.08^{+0.03}_{-0.10}$
       & $-0.04^{+0.07+0.10}_{-0.10-0.12}$
       & $-0.52\pm0.58$
      & $7.8^{+4.5}_{-2.0}$
      & $48.9^{+15.8+5.8}_{-13.7-12.5}$
      & $-64\pm48$
      \\
 $\eta' K_S$
       & $0.01^{+0.01}_{-0.02}$
       & $0.00^{+0.01+0.00}_{-0.02-0.00}$
       & $-0.19\pm0.09$
      & $1.7^{+0.4}_{-0.3}$
      & $2.1^{+0.2+0.1}_{-0.5-0.4}$
      & $7\pm7$
      \\
 $\eta K_S$
       & $0.07^{+0.03}_{-0.03}$
       & $0.07^{+0.03+0.00}_{-0.03-0.01}$
       & $-$
      & $-5.7^{+2.0}_{-5.5}$
      & $-3.9^{+1.8+2.5}_{-5.0-1.6}$
      & $-$
      \\
 $\pi^0K_S$
       & $0.06^{+0.03}_{-0.03}$
       & $0.04^{+0.01+0.02}_{-0.02-0.02}$
       & $-0.38\pm0.26$
      & $-3.2^{+1.1}_{-2.3}$
      & $3.7^{+1.9+1.7}_{-1.6-1.7}$
      & $2\pm13$
      \\
 \hline
\end{tabular}
\end{center}
\end{table*}

It is instructive to understand the size and sign of $\Delta {\cal
S}_f$ in the QCDF approach~\cite{QCDF}, for example.  Recall that
$\Delta S_f$ is proportional to the real part of $A^u_f/A^c_f$. We
follow \cite{QCDF} to denote a complex number $x$ by $[x]$ if
${\rm Re}(x)>0$. In QCDF the dominant contributions to
$A^u_f/A^c_f$ are basically given by~\cite{QCDF,BNb}
 \begin{eqnarray}
 \label{eq:QCDF}
 \frac{A^u_{\phi K_S}}{A^c_{\phi K_S}}
 &\!\!\!\!\sim&\frac{[-(a^u_4+r^\phi_\chi a^u_6)]}{[-(a^c_4+r^\phi_\chi a^c_6)]}\sim\frac{[-P^u]}{[-P^c]},
 \nonumber\\
 \frac{A^u_{\omega K_S}}{A^c_{\omega K_S}}
 &\!\!\!\!\sim&\frac{+[a^u_4-r^\phi_\chi a^u_6]+[a^u_2 R]}{+[a^c_4-r^\phi_\chi a^c_6]}\sim\frac{+[P^u]+[C]}{+[P^c]},
 \nonumber\\
 \frac{A^u_{\rho K_S}}{A^c_{\rho K_S}}
 &\!\!\!\!\sim&\frac{-[a^u_4-r^\phi_\chi a^u_6]+[a^u_2 R]}{-[a^c_4-r^\phi_\chi a^c_6]}\sim\frac{-[P^u]+[C]}{-[P^c]},
 \\
 \frac{A^u_{\pi^0 K_S}}{A^c_{\pi^0 K_S}}
 &\!\!\!\!\sim&\frac{[-(a^u_4+r^\phi_K a^u_6)]+[a^u_2 R']}{[-(a^c_4+r^\phi_K a^c_6)]}\sim\frac{[-P^u]+[C]}{[-P^c]},
 \nonumber\\
 \frac{A^u_{\eta' K_S}}{A^c_{\eta' K_S}}
 &\!\!\!\!\sim&\frac{-[-(a^u_4+r^\phi_K a^u_6)]+[a^u_2 R'']}{-[-(a^c_4+r^\phi_K a^c_6)]}\sim\frac{[-P^u]-[C]}{[-P^c]},
 \nonumber
 \label{eq:QCDF}
 \end{eqnarray}
where $a^p_i$ are effective Wilson coefficients~\footnote{In
general, we have Re$(a_2)>0$, Re$(a_6)<$Re$(a_4)<0$.}, $r_\chi=
O(1)$ are the chiral factors and $R^{(\prime,\prime\prime)}$ are
(real and positive) ratios of form factors and decay constants.

From Eq.(8), it is clear that $\Delta {\cal S}_f>0$ for $\phi
K_S$, $\omega K_S$, $\pi^0K_S$, since their Re$(A^u_f/A^c_f)$ can
only be positive. Furthermore, due to the cancellation between
$a_4$ and $r_\chi a_6$ in the $\omega K_S$ amplitude, the
corresponding penguin contribution is suppressed. This leads to a
large and positive $\Delta {\cal S}_{\omega K_S}$ as shown in
Table~I. For the cases of $\rho^0 K_S$ and $\eta' K_S$, there are
chances for $\Delta {\cal S}_f$ to be positive or negative. The
different signs in front of $[P]$ in $\rho^0 K_S$ and $\omega K_S$
are originated from the second term of the wave functions $(u\bar
u\pm d\bar d)/\sqrt2$ of $\omega$ and $\rho^0$ in the $\overline
B^0\to \omega$ and $\overline B^0\to\rho^0$ transitions,
respectively. The $[P]$ in $\rho^0 K_S$ is also suppressed as the
one in $\omega K_S$, resulting a negative $\Delta {\cal S}_{\rho^0
K_S}$. On the other hand, $[-P]$ in $\eta' K_S$ is not only
unsuppressed (no cancellation in the $a_4$ and $a_6$ terms), but,
in fact, is further enhanced due to the constructive interference
of various penguin amplitudes~\cite{BNa}. This enhancement is
responsible for the large $\eta' K_S$ rate~\cite{BNa} and also for
the small $\Delta {\cal S}_{\eta' K_S}$~\cite{QCDF,CCS2005}.

\section{FSI contributions to $\Delta {\cal S}_f$}

\begin{figure}[b]
\centering
\includegraphics[width=60mm]{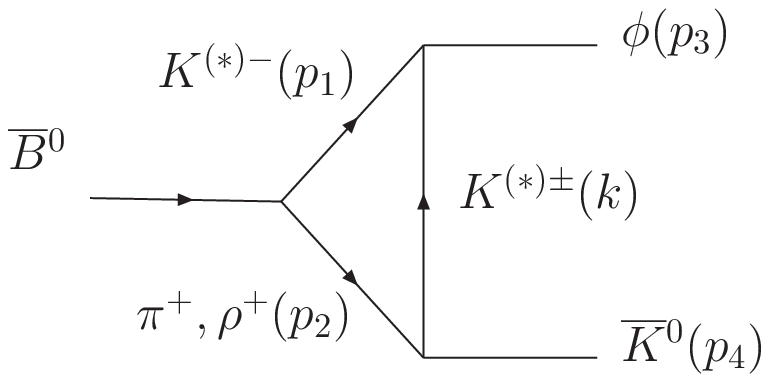}
\includegraphics[width=60mm]{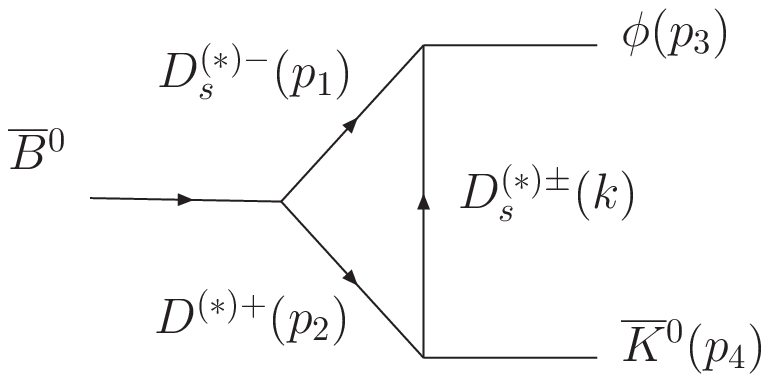}
\caption{Final-state rescattering contributions to the $\overline
B{}^0 \to\phi \overline K{}^0$ decay.} \label{example_figure}
\end{figure}

Evidence of direct CP violation in the decay $\overline B^0\to
K^-\pi^+$ is now established, while the combined BaBar and Belle
measurements of $\overline B^0\to\rho^\pm\pi^\mp$ imply a sizable
direct CP asymmetry in the $\rho^+\pi^-$ mode \cite{HFAG2006}. In
fact, direct CP asymmetries in these channels are much bigger than
expectations (of many people) and may be indicative of appreciable
LD rescattering effects, in general, in $B$ decays~\cite{CCS}. The
possibility of final-state interactions in bringing in the
possible tree pollution sources to ${\cal S}_f$ are considered
in~\cite{CCS2005}. Both $A_f^u$ and $A_f^c$ will receive
long-distance tree and penguin contributions from rescattering of
some intermediate states. In particular there may be some
dynamical enhancement of light $u$-quark loop. If tree
contributions to $A^u_f$ are sizable, then final-state
rescattering will have the potential of pushing $S_f$ away from
the naive expectation. Take the penguin-dominated decay $\overline
B^0\to \omega \overline K^0$ as an illustration. It can proceed
through the weak decay $\overline B^0\to K^{*-}\pi^+$ followed by
the rescattering $K^{*-}\pi^+\to \omega \overline K^0$. The tree
contribution to $\overline B^0\to K^{*-}\pi^+$, which is color
allowed, turns out to be comparable to the penguin one because of
the absence of the chiral enhancement characterized by the $a_6$
penguin term. Consequently, even within the framework of the SM,
final-state rescattering may provide a mechanism of tree pollution
to ${\cal S}_f$. By the same token, we note that although
$\overline B^0\to \phi \overline K^0$ is a pure penguin process at
short distances, it does receive tree contributions via
long-distance rescattering. Note that in addition to these
charmless final states contributions, there are also contributions
from charmful $D_s^{(*)} D^{(*)}$ final states, see Fig.~2. These
final-state rescatterings provide the long-distance $u$- and
$c$-penguin contributions.

An updated version of results in \cite{CCS2005} are shown in
Table~II. Several comments are in order. (i) $\phi K_S$ and $\eta'
K_S$ are the theoretical and experimental cleanest modes for
measuring $\sin2\beta_{\rm eff}$ in these penguin modes. The
constructive interference behavior of penguins in the $\eta' K_S$
mode is still hold in the LD case, resulting a tiny $\Delta {\cal
S}_{\eta'K_S}$. (ii) Tree pollutions in $\omega K_S$ and $\rho^0
K_S$ are diluted due to the LD $c$-penguin contributions.

It is found that LD tree contributions are in general not large
enough in producing sizable $\Delta {\cal S}_f$, since their
constributions are overwhelmed by LD $c$-penguin contributions
from $D_s^{(*)} D^{(*)}$ rescatterings. On the other hand, while
it may be possible to have a large $\Delta {\cal S}_f$ from
rescattering models that enhance the contributions from charmless
states, a sizable direct CP violation will also be generated.
Since direct CP violations are sensitive to strong phases
generated from FSI, these approaches will also give a sizable
direct CP violation at the same time when a large $\Delta {\cal
S}_f$ is produced. The present data on the $\phi K_S$ and $\eta'
K_S$ modes do not support large direct CP violations in these
modes. Consequently, it is unlikely that FSI will enlarge their
$\Delta {\cal S}_f$. In order to constrain or to refine these
calculations, it will be very useful to have more and better data
on direct CP violations.

\section{$\Delta {\cal S}_f$ in $KKK$ modes}

\begin{table}[b]
\caption{Mixing-induced and direct CP asymmetries $\Delta {\cal
S}_f$ (top) and ${\cal A}_f$ (in $\%$, bottom), respectively, in
$B^0\to K^+K^-K_S$ and $K_SK_SK_S$ decays. Results for
$(K^+K^-K_L)_{CP\pm}$ are identical to those for
$(K^+K^-K_S)_{CP\mp}$.} \label{tab:AS}
\begin{tabular}{|l| r r|}
\hline
 Final State & $\Delta {\cal S}_f$  & Expt.  \\
 \hline
 $(K^+K^-K_S)_{\phi K_S~{\rm excluded}}$
            & $0.03^{+0.08+0.02+0.00}_{-0.01-0.01-0.02}$
            & $-0.12^{+0.18}_{-0.17}$
            \\
 $(K^+K^-K_S)_{CP+}$
            & $0.05^{+0.11+0.04+0.00}_{-0.03-0.02-0.01}$
            &
            \\
 $(K^+K^-K_L)_{\phi K_L~{\rm excluded}}$
            & $0.03^{+0.08+0.02+0.00}_{-0.01-0.01-0.02}$
            & $-0.60\pm0.34$
            \\
 $K_SK_SK_S$
            & $0.02^{+0.00+0.00+0.01}_{-0.00-0.00-0.02}$
            & $0.19\pm0.23$
            \\
 $K_SK_SK_L$
            & $0.02^{+0.00+0.00+0.01}_{-0.00-0.00-0.02}$
            &
            \\
 \hline
  &${\cal A}_f(\%)$  &Expt. \\
 \hline
 $(K^+K^-K_S)_{\phi K_S~{\rm excluded}}$
            & $0.2^{+1.0+0.3+0.0}_{-0.1-0.3-0.0}$
            & $-8\pm10$
            \\
 $(K^+K^-K_S)_{CP+}$
            & $-0.1^{+0.7+0.2+0.0}_{-0.0-0.3-0.0}$
            &
            \\
 $(K^+K^-K_L)_{\phi K_L~{\rm excluded}}$
            & $0.2^{+1.0+0.3+0.0}_{-0.1-0.3-0.0}$
            & $-54\pm24$
            \\
 $K_SK_SK_S$
            & $0.7^{+0.0+0.0+0.1}_{-0.1-0.0-0.1}$
            & $31\pm17$
            \\
 $K_SK_SK_L$
            & $0.8^{+0.1+0.1+0.1}_{-0.3-0.1-0.1}$
            &
            \\
            \hline
 \end{tabular}
\end{table}

$\overline B {}^0\to K^+K^-K_{S,L}$ and $\overline B {}^0\to
K_SK_SK_S$ are penguin-dominated and pure penguin decays,
respectively. They are also used to extracted $\sin2\beta_{\rm
eff}$ with results shown in Fig.~1. Three-body modes are in
general more complicated than two-body modes. For example, while
the $K_SK_SK_S$ mode remains as a CP-even mode, the
$K^+K^-K_{S(L)}$ mode is not a CP-eigen state~\footnote{However,
it is found that $K^+K^-K_S$ is dominated by the CP-even part and
hence it is still useful in extracting $\sin2\beta_{\rm eff}$.}.
Furthermore, the mass spectra of these modes are in general
complicated and non-trivial.

A factorization approach is used to study these $KKK$
modes~\cite{CCSKKK}. In the factorization approach, the $\overline
B {}^0\to K^+K^-K_S$ amplitude, for example, basically consists of
two factorized terms: $\langle\overline B {}^0\to K_S\rangle\times
\langle0\to K^+K^-\rangle$ and $\langle\overline B {}^0\to K^+
K_S\rangle\times \langle 0\to K^-\rangle$, where $\langle A\to
B\rangle$ denotes a $A\to B$ transition matrix element. The
dominant contribution is from the $\langle\overline B {}^0\to
K_S\rangle\times \langle0\to K^+K^-\rangle$ term, which is a
penguin induced term, while the sub-leading $\langle\overline B
{}^0\to K^+ K_S\rangle\times \langle 0\to K^-\rangle$ term
contains both tree and penguin contributions. In fact, $\overline
B {}^0\to K^+ K_S$ transition is a $b\to u$ transition, which has
a color allowed tree contribution.

Results of CP asymmetries for these modes are given in Table~III.
The first uncertainty is from hadronic parameter in $\overline B
{}^0\to K^+ K_{S,L}$ transition in $K^+K^-K_{S,L}$ mode (and a
similar term in $K_SK_SK_S$ mode), the second uncertainty is from
other hadronic parameters, while the last uncertainty is from the
uncertainty in $\gamma$.

To study $\Delta {\cal S}_f$ and ${\cal A}_f$, it is crucial to
know the size of the $b\to u$ transition term ($A^u_f$). For the
pure-penguin $K_SK_SK_S$ mode, the smallness of $\Delta {\cal
S}_{K_SK_SK_S}$ and ${\cal A}_{K_SK_SK_S}$ can be easily
understood. For the $K^+K^-K_S$ mode, there is a $b\to u$
transition in the $\langle\overline B {}^0\to K^+
K_S\rangle\otimes \langle 0\to K^-\rangle$ term. It has the
potential of giving large tree pollution in $\Delta {\cal
S}_{K^+K^-K_S}$. It requires more efforts to study the size and
the impact of this term.

It is important to note that the $b\to u$ transition term in the
$K^+K^-K_S$ mode is not a CP self-conjugated term, since under a
CP conjugation, this term will be turned into a $\langle\overline
B {}^0\to K^- K_S\rangle\times \langle 0\to K^+\rangle$ term,
which is, however, missing in the original amplitude. Hence, this
term contributes to both CP-even and CP-odd configurations with
similar strength. Therefore, information in the CP-odd part can be
used to constrain its size and its impact on $\Delta {\cal S}_f$
and ${\cal A}_f$. Indeed, it is found
recently~\cite{Aubert:2005kd} that the CP-odd part is highly
dominated by $\phi K_S$, where other contributions (at
$m_{K^+K^-}\not= m_{\phi}$) are highly suppressed. Since the
$\langle\overline B {}^0\to K^+ K_S\rangle\times \langle 0\to
K^-\rangle$ term favors a large $m_{K^+K^-}$ region, which is
clearly separated from the $\phi$-resonance region, the result of
the CP-odd configuration strongly constrains the contribution from
this $b\to u$ transition term. Consequently, the tree pollution is
constrained and the $\Delta {\cal S}_{K^+K^-K_{S}}$ should not be
large. Note that results shown in Table~III were obtained without
fully incorporating these information. The first uncertainty in
Table~III will be reduced, if the CP-odd result is taken into
account. To further refine the results it will be very useful to
perform a detail Dalitz-plot analysis.

\begin{acknowledgments}
I am grateful to the organizers of FPCP2006 for inviting me to the
exciting conference and to Hai-Yang Cheng and  Amarjit Soni for
very fruitful collaboration.
\end{acknowledgments}

\bigskip 

\end{document}